# BIG BAD NUCLEOSYNTHESIS: A REVIEW

Lawrence M. Krauss[a] [*]

[a]Departments of Physics and Astronomy, Case Western Reserve University, Cleveland, OH 44106-7079

Big Bang Nucleosynthesis represents perhaps the first, and still perhaps the most powerful particle-astrophysics connection. As such, it should provide an example for other work in this area. I discuss the current status of standard model BBN predictions and constraints, and then argue that the issue of observational systematic uncertainties is the key feature limiting our ability to constrain theory with observation. Nevertheless, several very important constraints are currently obtainable. For example, assuming maximal systematic uncertainties in $^4$He, D, and $^7$Li we find a conservative upper limit of $\Omega_B \leq 0.16$ in order for BBN predictions to agree with observations. Equally significant, we find that BBN predictions are inconsistent unless $^4He$ abundance by mass is greater than 23.9%, or D $+^3$He estimates are incorrect. By contrast, unless systematic uncertainties are taken into account, the quoted $2\sigma$ observational upper limit on the primordial $^4He$ fraction is 23.8%.

## 1. THE POWER OF BBN

Your honor, there are two facts in this case which are not in dispute:

1.The predictions of standard BBN agree strikingly well with the inferred light element abundance estimates over many orders of magnitude
2. As a result there exist strong constraints on both non-standard and standard particle physics and cosmology

It is precisely these points which created the modern particle/astrophysics paradigm some 30 years or so ago. We are now living in the 1990's however, and it is appropriate to move beyond self-congratulation over the successes of the standard BBN model. In particular, there are two questions of particular relevance today:

1. What exactly *are* the constraints?
2. What precisely *are* the uncertainties?

It is the second question which is of greatest interest here. I believe that it points at what will, or at least what *should* be the trends in this subject today. As cosmology turns more and more into an empirical science the issue of understanding systematic uncertainties will become more and more central to utilizing data to constrain theory. Most important, if astroparticle constraints are to be believable, we must work hard not to over-interpret data, and derive constraints which may turn out later to be violated. As long as cosmology is dominated by *observations* and not *experiments*, the significance of systematic uncertainties should never be understated.

Most of this review will focus on recent work performed at CWRU to address these issues[1–3]. Before proceeding, In the spirit of this meeting, I want to briefly remind the reader how powerful BBN constraints can be for particle theory. Not only can they constrain in principle the physics associated with processes at MeV-scale temperatures, they can allow us to have a lever- arm which could in principle extend to the Planck scale! A few examples should suffice:

### 1.1. The Gravitino Problem

This problem has been around in one way or another as long as local supersymmetry has. The gravitino is the spin 3/2 supersymmetric partner of the graviton. In all such models of current interest, supersymmetry breaking is communicated to the observed particles by gravity, and the scale of supersymmetry breaking is manifested by a gravitino mass. Since supersymmetry fixes the gravitino couplings to matter to be iden-tical to the graviton couplings, the gravitino can decay into all ordinary particles with a coupling strength which is purely gravitational, and hence very weak. As a result, the graviton decay rate is fixed to be of order $\Gamma \approx (M^3/M_{pl}^2)$ where M is the gravitino mass. This implies that the graviton lifetime is generally much longer than O(1 sec) unless $M \geq (O10TeV)$. Because the gravitino is coupled to all particles, its decays will produce particles which can in turn photofission deuterium and helium, and thus destroy the good agreement of BBN predictions with observations. This problem can be resolved by diluting primordial gravitinos by inflation, but stringent requirements on reheating scales result (eg. [4,5]).

[*]CWRU-P1-95. To appear in Nucl. Phys. B, Proceedings and Supplements. Based on invited lectures at Trends in Astroparticle Physics, Stockholm, Sweden Sept 1994 and From the Weak Scale to the Planck Scale, Warsaw, Poland Sept 1994. Research supported in part by the DOE.

**1.2.** *Another Technicolor Problem*

While technicolor theories are beset with problems, they do have the virtue of being the only theories to attempt to fully dynamically address the problems of the observed quark and lepton mass matrices. If technicolor proports to be a theory of mass, it must also address, at some point, the possibility of neutrino masses. This is normally a difficulty for technicolor models, because if a see-saw type mechanism is utilized to produce Majorana masses, the smallness of the technicolor, or extended technicolor mass scale implies that the light neutrinos should not be very light. One way around this is to postulate light Dirac masses, which can be achieved by postulating symmetries which protect the observed neutrinos from heavy Dirac mass terms which give masses to techni-neutrinos. However, perhaps the most stringent constraint on such model building comes from BBN. The extra-right handed neutrinos are effectively sterile. Nevertheless, their interactions with their left-handed partners, resulting from extended techicolor interactions, are sufficient to populate them in the early universe at an unacceptable level unless the ETC scale is in excess of a few TeV [27]. This severely constrains such models.

**1.3.** *Neutrinos*

BBN is famous for constraining neutrinos. As I shall describe, this situation is getting even more interesting. We are on the verge of being able to address whether even completely "sterile" light neutrinos are allowed by BBN. In a related vein, even a singlet Majoran may be in trouble.

**1.4.** *Dark Matter*

BBN provides the strongest evidence available that the universe is not closed by Baryons. In turn, BBN constraints also provide strong motivation for the possibility that the inferred dark matter dominating galactic dynamics is non-baryonic. As I shall describe, results allow the possibility in principle to convincingly resolve this issue. At the same time, recent results on the baryon fraction of dense clusters provides a potential confrontation with the idea that the universe is flat, unless systematic uncertainties forced BBN constraints to be relaxed. With this in mind, several groups have recently investigated how large the baryon abundance of the universe may be pushed in standard BBN models.

## 2. A BRIEF PRIMER

Since we shall be interested here in how BBN predictions depend on several fundamental microphysical and cosmological parameters, including the number of light neutrinos, the neutron lifetime, and the baryon density of the universe, I thought I would spend a few paragraphs outlining why this is the case.

First, the heart of BBN calculations is the equation which is at the heart of much of particle-astrophysics, the Boltzmann equation in an expanding universe. The number density, $n$, of any particle species evolves as

$$dn/dt = - <\sigma v>_a n^2 + \Sigma_i <\sigma v>_p^i X_i^2 - 3n\dot{R}/R$$

Here, the first term on the RHS represents the thermally averaged annihilation cross section, the second term represents the sum over thermally averaged production cross sections, and the final term represents dilution due to the expansion of the universe.

This equation permeates BBN considerations. It is responsible for (a) determining the remnant neutron number density, (b) determining the remnant abundances of all light elements, and (c) determining the number density of light neutri-nos and other exotic particle species which might affect the expansion rate of the universe during BBN. The key point is that once the last term in the equation dominates the two preceeding terms, the number density of a given particle species begins to depart from its thermal equilibrium value. Massless particles continue to keep their thermal equilibrium form, but the temperature characterising the distribution need not be the same as the temperature of the remaining particles in thermal equilibrium if entropy subsequently gets dumped into the radiation gas (Hence, some particles can have an abundance which is equivalent to a fraction of a neutrino in thermal equilibrium). Massive particles, such as neutrons, have a distribution which freezes out, so that their number density is suppressed by a Boltzmann factor characteristic of the temperature at freeze-out, rather than continuing to fall with decreasing temperature. Finally, composite particles such as helium nuclei will not be produced with their thermal equilibrium value until the abundance of reactants involved in their production (i.e. $X_i$) reaches a critical value.

This completely explains the dependence of the residual $^4He$ abundance produced during BBN on the fundamental parameters described briefly in the subsections below. Since one of the purposes of this review will be to demonstrate that almost all the BBN action today revolves around $^4He$, this will suffice for our needs.

### 2.1. *The neutron lifetime*

The neutron lifetime is one way of parametrizing the strength of the weak interactions which interconvert neutrons and protons. The longer the neutron lifetime, the weaker these reactions are. As a result, the longer the neutrino lifetime, the earlier the weak interactions which keep neutrons and protons in thermal equilibrium decouples. As a result, more remnant neutrons will be available to partake in BBN reactions. In turn, more helium can be produced during BBN.

### 2.2. *The number of neutrinos*

The expansion rate of the universe is directly proportional to the density of the radiation gas, which is directly proportional to the number of species in the radiation gas. As this number increases by an amount which is equivalent to one extra neutrino helicity degree of freedom, the expansion rate increases by a fixed amount. This in turn implies that the weak interactions decouple at a slightly higher temperature, which in turn results in more primordial $^4He$ being produced. One of the recent results I shall describe is a new derivation of the relation between primordial helium and neutrino number.

### 2.3. *The baryon density*

Primordial $^4He$ cannot form until enough deuterium forms so that helium production reactions can compete with the expansion rate of the universe. The earlier this happens, the more efficient is helium production (in part because neutrons are decaying slowly until they get bound in nuclei). The greater the primordial baryon density, the greater the density of neutrons and protons. The greater the density of neutrons and protons, the larger the production rate of deuterium. The larger the production rate of deuterium, the earlier a critical density of deuterium nuclei forms which can result in rapid production of helium. Thus, the greater the baryon density the larger the primordial helium fraction produced.

## 3. BBN: RECENT DEVELOPMENTS

The remarkable agreement of the predicted primordial light element abundances and those inferred from present observations yields some of the strongest evidence in favor of a homogeneous FRW Big Bang cosmology. Because of this, significant efforts have taken place over 20 years to refine BBN predictions, and the related observational constraints. Several factors have contributed to the maturing of this field, including the incorporation of elements beyond $^4$He in comparison between theory and observation[6], and more recently: an updated BBN code [7], a more accurate measured neutron half life[8], new estimates of the actual primordial $^4$He , $D +^3 He$, and $^7Li$ abundances [9,10], and finally the determination of BBN uncertainties via Monte Carlo analysis [11]. All of these, when combined together[12], yield a consistent and strongly con-strained picture of homogeneous BBN.

We recently returned to re-analyze BBN constraints initially motivated by three factors: new measurements of several BBN reactions, the development of an improved BBN code, and finally the realization that a correct statistical determination of BBN predictions should include correlations between the different elemental abundances. Each serves to further restrict the allowed range of the relevant cosmological observables $\Omega_B$ and $N_\nu$. Of course, statistically determined uncertainties are not the major factor limiting our ability to use BBN to constrain fundamental parameters. As we shall see, systematic uncertainties in the inferred light element abundances are generally much larger, and must be properly accounted for if we are to conservatively compare predictions with observations. In this first section I outline the details of our effort to properly update and account for BBN statistical uncertainties, and leave the discussion of systematics to a later section.

### 3.1. New BBN Reaction Rates:

By far the most accurately measured BBN input parameter is the neutron half-life, which governs the strength of the weak interaction which interconverts neutrons and protons. Since this effectively determines the abundance of free neutrons at the onset of BBN, it is crucial in determining the remnant abundance of $^4$He. With the advent of neutron trapping, the uncertainty in the neutron half-life quickly dropped to less than 0.5% by 1990. Nevertheless, it is the uncertainty in this parameter that governs the uncertainty in the predicted $^4$He abundance. The world average for the neutron half-life is now $\tau_N = 887 \pm 2 sec$ [8] (Note: This is a recent update from the earlier value of $\tau_N = 889 \pm 2.1 sec$. Unless otherwise stated, the new value is used in the tables and formulae presented here and in our most recent work [3], which thus updates values found in some of our earlier work[1,2]). This has an uncertainty almost twice as small as that used in vaious earlier analyses.

We also updated the rate $^7Be + p \to \gamma + ^8B$ [13] 20% which we thought might be significant at high values of $\eta_{10}$ (defined by $\Omega_B = .0036 h^{-2} (T/2.726)^3 \eta_{10} \times 10^{10}$, where $T$ is the microwave background temperature today, and $h$ defines the Hubble parameter $H = 100h$ km/(Mpc sec)), but is not. Other than these two new rates we used the reaction rates and uncertainties from [12].

### 3.2. New BBN Monte Carlo:

Because of the new importance of small corrections to the $^4$He abundance when comparing BBN predictions and observations, increased attention has been paid recently to effects which may alter this abundance at the 1% level or less. In the BBN code several such effects were incorporated, resulting in an $\eta_{10}$-independent correction of $+.0006$ to the lowest order value of $Y_p$ (the $^4$He mass fraction). This is a change of $+.0031$ compared to the value used in previous published analyses[9,11].

In the present code, more than half of the new correction is due to finer integration of the nuclear abundances. Making the time-step in the code short enough that different Runge-Kutta drivers result in the same number for the $^4$He abundance produces a nearly $\eta_{10}$ independent change in $Y_p$ of $+.0017$ [15]. Residual numerical uncertainties are small [15,16]. The other major change is the inclusion of $M_N^{-1}$ effects[17]. Seckel showed that the effects on the weak rates due to nucleon recoil, weak magnetism, thermal motion of the nucleon target and time dilation of the neutron lifetime combine to increase $Y_p$ by $\sim .0012$. Also included in the correction is a small increase of .0002 in $Y_p$ from momentum dependent neutrino decoupling [18,19].

Finally, we utilized a Monte Carlo procedure in order to incorporate existing uncertainties and determine confidence limits on parameters. Such a procedure was first carried out[11] with BBN reaction rates chosen from a (temperature-independent) distribution based on then existing experimental uncertainties. This procedure was further refined[12] by updating experimental uncertainties and using temperature dependent uncertainties. Here we utilized the nuclear reaction rate uncertainties in [12] (including the temperature dependent uncertainties for $^3$He$(\alpha, \gamma)^7$Be and $^3$H$(\alpha, \gamma)^7$Li) except for the reactions we updated. Each reaction rate was determined using a

Gaussian distributed random variable centered on unity, with a $1-\sigma$ width based on that quoted in [12]. For the rates without temperature- dependent uncertainties this number was used as a multiplier throughout the integration. For the two rates with temperature-dependent uncertainties the original uniformly distributed random number was saved and mapped into a new gaussian distribution with the appropriate width at each time step.

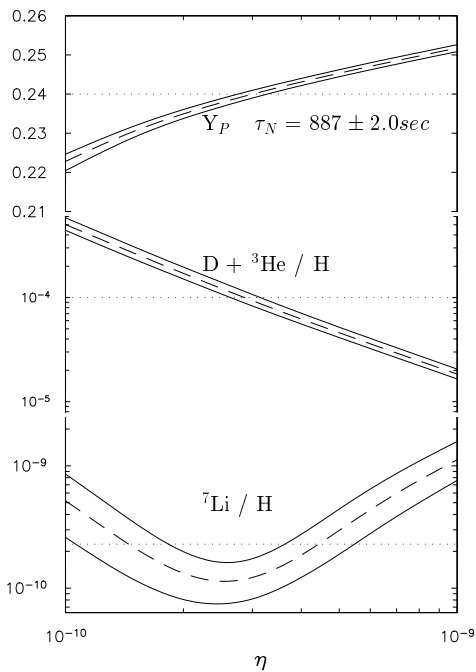

Figure 1. BBN Monte Carlo predictions as a function of $\eta_{10}$. Shown are symmetric 95% confidence limits on each elemental abundance. Also shown are claimed upper limits inferred from observation.

The results of the updated BBN Monte Carlo analysis are displayed in figure 1, where the symmetric 95% confidence level predictions for each elemental abundance are plotted. Also shown are previously claimed observational upper limits for each of the light elements [9,10,12]. *In the first instance we shall utilize these limits in order to assess how BBN constraints have evolved based on our re-analysis, and after this we shall consider how to account for systematic uncertainties.* Figure 1 also allows one to assess the significance of the corrections we used, in relation to the width of the 95% C.L. band for $Y_p$, which turns out to be $\sim$ .002. The total change in $Y_p$ of $\approx +.003$ from previous BBN analyses conspires with the reduced uncertainty in the neutron lifetime, which narrows the uncertainty in $Y_p$ and feeds into the uncertainties in the other light elements, to reduce the range where the predicted BBN abundances are consistent with the inferred primordial abundances.

### 3.3. Statistical Correlations Between Predicted Abundances:

While the introduction of a Monte Carlo procedure was an important step, the determination of limits on the allowed range of BBN parameters $\Omega_B$ and $N_\nu$ based on comparison of symmetric 95% confidence limits for single elemental abundances with observations, as has become the standard procedure, overestimates the allowed range. This is because the BBN reaction network ties together all reactions, so that the predicted elemental abundances are not statistically independent. In addition, the use of symmetric confidence limits is too conservative. Addressing these factors is a central feature of our work.

Figure 2 displays the locus of predicted values for the fractions $Y_p$ and D $+^3$He/H for 1000 BBN models generated from the distributions described above for $\eta_{10} = 2.71$ (figure a) and $\eta_{10} = 3.08$ (figure b). Also shown is the $\chi^2 = 4$ joint confidence level contour derived from this distribution, in a Gaussian approximation, calculating variances and covariances in the standard manner. The horizontal and vertical tangents to this contour correspond to the individual symmetric $\pm 2\sigma$ limits on Gaussianly distributed $x$ and $y$ variables. As is evident from the figure, and as is also well known on the basis of analytical arguments, there is a strong anti-correlation between $Y_p$ and the remnant D $+^3$He abundance (the normalized covariance ranges from -0.7 to -0.4 in the $\eta_{10}$ range of interest). Thus, those models where $^4$He is lower than the mean, and which therefore may be allowed by an upper bound of 24% on $Y_p$, will also generally produce a larger remnant D+$^3$He/H abundance, which can be in conflict with the bound on this combination of $10^{-4}$ [24]. This will have the effect of reducing the parameter space which is consistent with both limits.

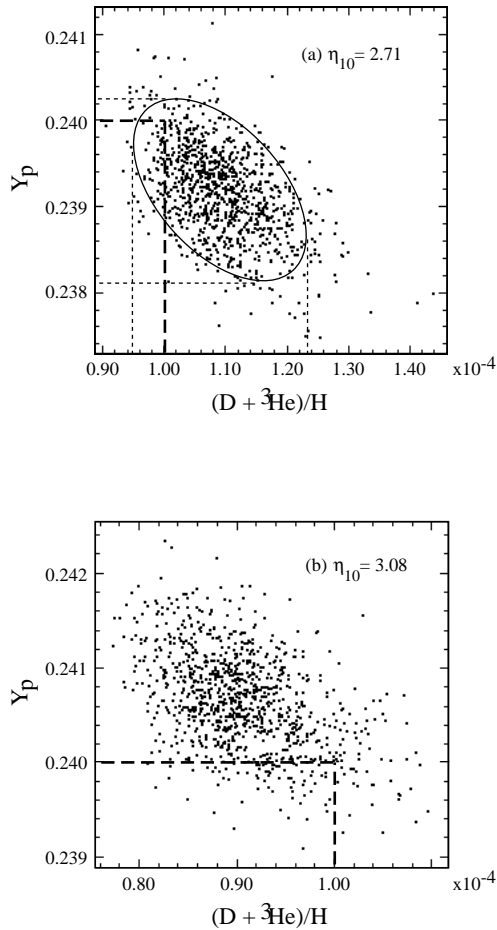

Figure 2. Monte Carlo BBN predictions for $Y_p$ vs $D+^3He$ and allowed range for (a) $\eta_{10} = 2.71$, and (b) $\eta_{10} = 3.08$ (using 1992 neutron half life value). In (a) a Gaussian contour with $\pm 2\sigma$ limits on each individual variable is also shown.

Because our Monte Carlo generates the actual distribution of abundances, Gaussian or not, we determined a 95% confidence limit on the allowed range of $\eta_{10}(N_\nu)$ by requiring that at least 50 models out of 1000 lie within the joint range bounded by both the $^4$He and D +$^3$He upper limits, as shown in figure 2. This is to be compared with the procedure which one would follow without considering joint probability distributions. In this case, one would simply check whether 50 models lie *either* to the left of the D +$^3$He constraint for low $\eta_{10}$ (figure a), or below the $^4$He constraint for high $\eta_{10}$ (figure b). This is clearly a looser constraint than that obtained using the joint distribution. Finally, the procedure which has been used to-date, which is to check whether the symmetric $2\sigma$ confidence limit (i.e. when 25 models exceed either bound) for a single elemental abundance crosses into the allowed region gives even a looser constraint, as can be seen in figure 2a.

In table 1 we display these results. Here we show the 95% confidence limits on $\eta_{10}$, both as we have defined them above and also using the looser procedures which ignore correlations. Note that the constraints tighten dramatically as the number of effective light neutrino species, $N_\nu$ is increased.

### 3.4. Implications and Caveats:

The above constraints on $\eta_{10}$ and $N_\nu$, taken at face value, assuming a $Y_p$ upper bound of 24% and an D+$^3$He/H upper limit of $10^{-4}$, would have significant implications for cosmology, dark matter and particle physics. The limit on $\eta_{10}$ corresponds to the limit $0.015 \leq \Omega_B \leq 0.070$. (Assuming $0.4 \leq h \leq 0.8$, as is required by direct measurements and limits on the age of the universe.) Thus, *if the previously quoted observational upper limits on the $Y_p$ and D+$^3$He/H are used directly*, homogeneous BBN would imply:

(a) The upper limit on $\Omega_B$ would be marginally incompatible with even the value of $\approx 0.1$ inferred from rotation curves of individual galaxies, further suggesting the need for non-baryonic dark matter in these systems.

(b) The bound on the number of effective light degrees of freedom during nucleosynthesis is *very severe*, corresponding to less than 0.04 extra light neutrinos with the 1992 neutron half life and .07 extra neutrinos with the new world average (see later section). This is perhaps the most worri-some bound of all because it argues against *any* Dirac mass for a neutrino without some significant extension of the standard model. This is because even a light "sterile" right handed component whose interactions freeze out about 300 GeV will still contribute in excess of 0.047 extra neutrinos during BBN [20] without extra particles whose annihilations can further suppress its abundance compared to its original thermal abundance. It can have further implications for a variety of kinds of hot or cold dark matter. For example, new light scalars would be ruled out unless they decouple above the electroweak scale, as would be any significant population of supersymmetric particles during BBN. Even allowing 0.047 extra light neutrinos, the upper limit on a Dirac mass would be reduced to ≈ 5 keV [25,26]. A $\nu_\tau$ mass greater than 0.5 MeV with lifetime exceeding 1 sec. would also be ruled out due to its effect on the expansion rate during BBN[21,23]. Also, neutrino interactions induced by extended technicolor at scales less than $O(100)$ TeV are ruled out [27], and sterile right handed neutrinos [28] would be ruled out as warm dark matter as the lower limit on their mass would now be $O(1keV)$.
(c) *The primordial $^4$He mass fraction would have to be greater than 23.9 % for consistency with D+$^3$He constraints!* [3] As we shall describe, this generally exceeds the $2\sigma$ upper limit on the primordial $^4$He abundance inferred without accounting for possible systematic uncertainties in the analysis.

Table 1
Correlations & $\eta_{10}$ Limits (1992 $\tau_{neutron}$ value)

| 95% C.L. $\eta_{10} range$ | $N_\nu$ | | | |
|---|---|---|---|---|
| | 3.0 | 3.025 | 3.04 | 3.05 |
| with corr. | 2.69 ↔ 3.12 | 2.75 ↔ 2.98 | 2.83 ↔ 2.89 | ∅ |
| no corr. | 2.65 ↔ 3.14 | 2.65 ↔ 3.04 | 2.69 ↔ 2.99 | 2.69 ↔ 2.95 |
| sym. no corr. | 2.62 ↔ 3.17 | 2.63 ↔ 3.10 | 2.65 ↔ 3.03 | 2.66 ↔ 3.00 |

These constraints are so stringent that they cry out for a consideration of uncertainties in the light element abundance estimates. Indeed, as we shall next discuss, in spite of the considerable effort devised above to accomodate statistical uncertainties in the predictions, by far the largest and most significant uncertainties in the comparison of BBN predictions and observations come from the latter. Moreover, these uncertainties are systematic and not statistical. Accomodating them in a BBN analysis will be the subject of the rest of this review.

## 4. ABUNDANCE ESTIMATES AND SYSTEMATIC UNCERTAINTIES

As we have just indicated, the weakest link in a BBN analysis involves the assumed light element abundances. Estimates of $^4$He , for example [30–32] are mostly indirect, and subject to large systematic uncertainties, which may also be important for the other abundance estimates. As a result, our refined BBN analysis described above suggests the need for revision of the light element abundance estimates inferred from observation at least as much as it argues for or against new nonstandard physics.

One of the most worrisome aspects of the present constraints is the fact that D+$^3$He provides a lower limit on $Y_p$ which is uncomfortably close the previously claimed upper limit. There are obviously two ways out of this dilemma: either the D+$^3$He limit is increased, which would allow smaller values of $Y_p$ to be consistent, or the observational upper limit on $Y_p$ is increased. Both possibilities have recently been discussed as I shall describe below. In addition, recently, several groups have assessed more carefully the systematic uncertainties present particularly in the primordial $^4$He abundance estimates[36,37,42], and have quoted various new upper limits on cosmological parameters based on their assessments. It is very clear, based in part on the differing estimates, that it is quite difficult at the present time to get an accurate handle on these uncertainties.

Because of this, and because we could utilize the full statistical machinery we previously developed when comparing predictions to "observations", we felt it would be useful to prepare a comprehensive table of constraints on $N_\nu$ and $\Omega_B$ for a relatively complete range of different assumptions about light element abundances. This allowed us to explore the role of different estimates in the constraints, as well as the effect of correlations as the light element abundance estimates vary. In addition, it allows us to address several points which we feel are important to consider when deriving cosmological constraints using BBN predictions. Finally, this analysis leads to new simple relations between the light element abundances and limits on cosmological parameters such as the number of neutrinos, $N_\nu$, and the baryon to photon ratio $\eta_{10}$.

Before proceeding, one might wonder whether if systematic uncertainties in the inferred primordial element abundances are dominant, one need concern oneself with the proper handling of statistics in the predicted range. Thankfully, there are two factors which make the comparison of predictions and observations less ambiguous in the case of BBN:

(1) Because the allowed range in the observationally inferred abundances is much larger than the uncertainty in the predicted abundances, any constraint one deduces by comparing the two depends merely on the upper *or* lower observational limit for each individual element, and not only both at the same time. Thus, one is not so much interested in the entire distribution of allowed abundances as one is in one extremum of this distribution.

(2) Systematic uncertainties dominate for the observations, while statistical uncertainties dominate for the predictions.

Both of these factors suggest that a conservative but still well defined approach involves setting *strict* upper limits on $Y_p$, D+$^3$He, and $^7$Li, and a lower limit on D, which incorporate the widest range of reasonably accepted systematic uncertainties. Determining what is reasonable in this sense is of course where most of the "art" lies. Nevertheless, once such limits are set and treated as strict bounds, then one can compare correlated predictions with these limits in a well defined way. In this way one replaces the ambiguity of properly treating the distribution of observational estimates with what in the worst case may be a somewhat arbitrary determination of the extreme allowed observational values.

Clearly all the power, or lack thereof, in this procedure lies in the judicious choice of observational upper or lower limits. Because of our concern about the ability at present to prescribe such limits I consider a variety of possibilities here. Once one does choose such a set, however, it is inconsistent not to use all of it throughout in deriving constraints. If one uses one observational upper limit for $Y_p$, for example, to derive constraints on the number of neutrinos, but does not use it when deriving bounds on the baryon density, then probably one has not chosen a sufficiently conservative bound on $Y_p$ in the former analysis. It has been argued that a weak, logarithmic, dependence of $Y_p$ on $\eta_{10}$ invalidates its use in deriving bounds on the latter quantity. Not only can this argument be somewhat misrepresentative for an interesting range of $Y_p$ values, but until $Y_p$ exceeds statistically derived upper limits by a large amount, it can continue to play a signifcant role in bounding $\eta_{10}$ from above.

### 4.1. Abundance Estimate Uncertainties: The Range

(a) $^4$He: By correlating $^4$He abundances with metallicity for various heavy elements including O,N and C, in low-metallicity HII regions one can attempt to derive a "primordial" abundance defined as the intercept for zero metallicity. This can be determined by a best fit technique, assuming some linear or quadratic correlation between elemental abundances (i.e. see [22,44–46]). The statistical errors associated with such fits are now small. Best fit values obtained typically range from .228-.232, with statistical "$1\sigma$" errors on the order of .003-.005. This argument yields the upper limit of .24 [22] which has been oft quoted in the literature. Recently this number has begun to drift upwards slightly. New observations of HII regions in metal poor galaxies have tended to increase the statistically derived zero intercept value of $Y_p$ by perhaps .005 (i.e. [36,47]). In addition, the recognition that systematic, and not statistical uncertainties may dominate any such fit has become more widespread recently. The key systematic uncertainty which interferes with this procedure is the uncertainty in the $^4$He abundance determined for each individual system, based on uncertainties in modelling HII regions, ionization, etc used to translate observed line strengths into mass fractions. Many observational factors come into play here (see [48] for a discussion of observational uncertainties), and people have argued that one should add an extra systematic uncertainty of anywhere from .005-.015 to the above estimate. Clearly thus, one should examine implications of $^4$He abundances in the range .24-.25. We shall show that for $Y_p$ above .25; (a) $^4$He becomes unimportant for bounding $\eta_{10}$, and (b) the effect on bounds on $N_\nu$ can be obtained by straightforward extrapolation from the data obtained for the range .24-.25.

(b) $^7$Li: It is by now generally accepted that the primordial abundance of $^7$Li is closer to the Spite Pop II plateau than the Pop I plateau. Nevertheless, even if one attempts to fit the primordial abundance by fitting evolutionary models to the Pop II data points[10], assuming no depletion, one still finds a $2\sigma$ upper limit as large as $2.3 \times 10^{-10}$. The role of rotationally induced depletion is still controversial. It is clear some such depletion is expected, and can be allowed for[41], but observations of $^6$Li, which is more easily depleted, put limits on the amount of $^7$Li depletion which can be allowed. We assumed an extreme factor of 2 depletion as allowable, and thus we explore how cosmological bounds are affected by a $^7$Li upper limit as large as $\approx 5 \times 10^{-10}$.

(c) D and D+$^3$He: The situation regarding this combination has become quite interesting recently. There has been a new claimed observation[33], of deuterium in a primordial gas cloud, at a level (D+$^3$He)/H $= 1.9 - 2.5 \times 10^{-4}$. It has long been argued that any present measurement of $D$ provides a lower limit on its primordial abundance because $D$ is so fragile that it is easily destroyed in stars. Also, because the predicted BBN abundance falls monotonically with increasing baryon density, a lower limit on deuterium thus places a reliable upper limit on the baryon density of the universe. Previously quoted solar system abundance estimates of $10^{-5}$ led to a firm upper bound on $\eta_{10} < 8$ which clearly established that baryons could not close the universe. The Songaila et al. observation, an order of magnitude larger, is also a factor of two greater than the previous upper limit on the combination D+$^3$He. As a result, this would allow smaller values of $\eta_{10}$, which would in turn allow a smaller value of $Y_p$, although the upper limit on $\Omega_B$ one might derive would be much more severe. We have recently explored the implications of this possibility in some depth [2]. In particular, we have shown that this result, if upheld, would change the way we combine elemental abundance limits to get constraints on cosmology and particle physics. Partly for this reason, in the analysis to be described below the solar system D abundance of $2 \times 10^{-5}$ is taken as a firm lower bound on D, and the previously quoted upper limit of $10^{-4}$ is utilized as an upper limit on D+$^3$He [43]. In the first place, it must be stressed that the Songaila et al result is still preliminary, and could easily be due to interloping hydrogen clouds along the line of sight of the system being observed. Also, other arguments based on galactic evolution and the pre-solar D+$^3$He abundance are in apparent contradiction with the result. In any case, if the Songaila et al result were to be confirmed, it would require a discrete and dramatic shift in the entire BBN analysis, as we have described elsewhere. Thus, while one should keep in mind that such a parallel constraint space is a possibility, it is sufficiently "non-standard" so that it should be considered independent of the more generic systematic uncertainties we wish to concentrate on here.

## 4.2. Results and Tables

Tables 2-4 give our key results, adn the following description of them is taken from [3]. The data were obtained using 1000 Monte Carlo BBN runs at each value of $\eta_{10}$, with nuclear reaction rate input parameters chosen as Gaussian random variables with appropriate widths (see [1] for details) . In each case the number of runs which resulted in abundances which satisfied the joint constraints obtained by using combinations of the upper limits on $^4$He, $^7$Li, and D+$^3$He or the lower limit on D was determined. Limits on parameters were determined by varying these until less than 50 runs out of 1000 (up to $\sqrt{N}$ statistical fluctuations) satisfied all of the constraints.

Table 2 displays the upper limit on $N_\nu$ for various values of $Y_p$. As is shown, this was governed by the combination of $^4$He and D+$^3$He upper limits. Shown in the table are the number of acceptable runs out of 1000 when the two elemental bounds are considered separately and together, for an $\eta_{10}$ range which was found to maximize the number of acceptable models. Throughout the $Y_p^{max}$ region from .24 to .25, both the $Y_p$ and D+$^3$He limits play a roughly equal role in determining the maximum value of $N_\nu$. We are able to find a remarkably good analytical fit for the maximum value of $N_\nu$ as a function of $Y_p$ as follows:

$$N_\nu^{max} = 3.07 + 74.07(Y_p^{max} - .240) \quad (1)$$

The linearity of this relation is striking over the whole region from .24 to .25 in spite of the interplay between the two different limits in determining the constraint. Note also that this relation differs from that quoted in [22] between $Y_p$ and $N_\nu$ in that the slope we find is about 13% less steep than that quoted there. The two formulae are not strictly equivalent in that the one presented in [22] presented the best fit value of $Y_p$ determined in terms of $N_\nu$, while the present formula gives a relation between the maximum allowed values of these parameters, based on limits on the *combination* $Y_p$ and D+$^3$He, and on the width of the predicted distribution. In this sense, eq. (1) is the appropriate relation to utilize when relating bounds on $Y_p$ to bounds on $N_\nu$.

Tables 3 and 4, which display the upper bounds on $\eta_{10}$, are perhaps even more enlightening. They demonstrate the sensitivity of the upper limit on $\eta_{10}$ and hence $\Omega_{baryon}$ to the various other elemental upper limits as $Y_p$ is varied. Several features of the data are striking. First, note that $^4$He completely dominates in the determination of the upper limit on $\eta_{10}$ until $Y_p$ =.245, even for the most stringent chosen upper limit on $^7$Li. If this limit on $^7$Li is relaxed, then $^4$He dominates as long as the upper limit on $Y_p \leq$.248! Also note that the "turn on" in significance of the $^7$Li contribution to the constrain is somewhat more gradual than the "turn off" of the $^4$He constraint. The former turns on over a range of $\eta_{10}$ of about 2, while the latter turns off over a range of about 1-1.5. This gives one some idea of the size of the error introduced in determining upper bounds by using only either element alone, rather than the combination. Next, for a $Y_p$ upper limit which exceeds .248, the lower bound on D begins to become important. It quickly turns on in significance so that by the time the upper limit on $Y_p$ is increased to .25, $^4$He essentially no longer plays a role in bounding $\eta_{10}$. Finally, note that both the relaxed bound on $^7$Li and the D bound converge in significance at about the same time, so that for $\eta_{10} > 7.25$, both constraints are significantly violated. This implies a "safe" upper limit on $\eta_{10}$ at this level, which corresponds to an upper bound $\Omega_{baryon} \leq .163$, assuming a Hubble constant in excess of 40 km/sec/Mpc. We again stress that a value this large is only allowed if $Y_p$ exceeds .250. If, for example, $Y_p \leq .245$, then the upper bound on $\Omega_{baryon}$ is essentially completely determined by $^4$He and is then at most 0.11. These limits may be compared to recent estimates of $\Omega_{baryon}$ based on X-ray determinations of the baryon fraction in clusters [49].

One final comment on the role of $Y_p$ in constraining $\eta_{10}$: It has been stressed that because of the logarithmic dependence of the former on the latter, that $Y_p$ cannot be effectively used to give a reliable upper bound on $\eta_{10}$. This is somewhat deceptive, however. We can compare how much more sensitive the bound on $\eta_{10}$ is to $Y_p$ than the bound on $N_\nu$ is by making a linear fit to the former relation and comparing it to (1). If we do this, we find first that the linear fit is quite good out to $Y_p$ as large as .245 (after which a quadratic fit remains good all the way out to .248, where the D and relaxed $^7$Li bounds begin to take over), and is given by

$$\eta_{10}^{max} \approx 3.22 + 354(Y_p^{max} - .240) \quad (2)$$

Seen in these terms, the $\eta_{10}$ upper limit is approximately 4.5 times more sensitive to the precise upper limit chosen for $Y_p$ than is the $N_\nu$ upper limit. Thus, while there is no doubt that varying the up-per limit on $Y_p$ has a more dramatic effect on the upper bound one might derive for $\eta_{10}$ than it does for constraining $N_\nu$, the quantitative nature of the relative sensitivities is perhaps displayed, for the relevant range of $Y_p$, by comparing the linear approximations presented here than by discussing logarithmic vs linear dependencies. More important, even recognizing the increased sensitivity of $\eta_{10}$ on $Y_p$, unless one is willing to accept the possibility of a rigid upper bound on $Y_p$ greater than .247, it is overly conservative to ignore $^4$He when deriving BBN bounds on $\eta_{10}$.

Table 2
$^4$He Abundance Estimates & $N_\nu$ limits

| $Y_p$ | $N_{\nu_{max}}$ | # allowed models: {$^4$He & [D+$^3$He]}($^4$He:D+$^3$He) | | | |
|---|---|---|---|---|---|
| | | $\eta_{10}$=2.75 | 2.80 | 2.85 | 2.90 |
| .240 | 3.07 | 40(603:148) | 52(429:254) | 46(268:376) | 38(170:534) |
| | | $\eta_{10}$=2.80 | 2.85 | 2.90 | 2.95 |
| .241 | 3.14 | 38(532:171) | 46(354:309) | 39(219:470) | 35(131:625) |
| .242 | 3.21 | 41(562:154) | 55(451:276) | 53(272:423) | 52(163:616) |
| .243 | 3.29 | 17(588:110) | 32(410:220) | 46(266:378) | 36(184:513) |
| .244 | 3.36 | 30(669:102) | 44(501:187) | 38(353:336) | 40(216:464) |
| | | $\eta_{10}$=2.85 | 2.90 | 2.95 | 3.00 |
| .245 | 3.43 | 50(598:173) | 68(449:296) | 64(308:427) | 54(173:586) |
| .247 | 3.59 | 27(635:84) | 30(480:184) | 47(338:306) | 39(185:488) |
| | | $\eta_{10}$=2.95 | 3.00 | 3.05 | 3.10 |
| .250 | 3.82 | 45(491:207) | 47(364:374) | 50(225:495) | 32(131:587) |

Table 3
$^4$He and $^7$Li Abundance Estimates & $\eta_{10}$ limits

| $Y_{p_{max}}$ | $\eta_{10_{max}}$ ($^7$Li$_{-10}$ < 2.3) | # allowed models: {$^4$He&$^7$Li} ($^4$He:$^7$Li) | $\eta_{10_{max}}$ ($^7$Li$_{-10}$ < 5) | # allowed models: {$^4$He&$^7$Li}($^4$He:$^7$Li) |
|---|---|---|---|---|
| .240 | 3.26 | 56 (60:998) | 3.26 | 56 (60:1000) |
| .241 | 3.55 | 45 (45:986) | 3.55 | 45 (45:1000) |
| .242 | 3.89 | 45 (47:905) | 3.89 | 47 (47:1000) |
| .243 | 4.26 | 50 (60:626) | 4.27 | 46 (46:1000) |
| .244 | 4.64 | 48 (92:296) | 4.71 | 49 (49:1000) |
| .245 | 5.01 | 45 (211:118) | 5.23 | 62 (62:984) |
| .246 | 5.23 | 51 (679:62) | 5.80 | 46 (50:810) |
| .247 | 5.25 | 52 (997:52) | 6.36 | 48 (80:500) |

Finally, it is in this analysis that we rederived the minimum value of $Y_p$ such that BBN predictions are consistent with observation. We explored the range of $\eta_{10}$ allowed at the 95% confidence level (i.e. 50 out of 1000 models) as the value of $Y_p^{max}$ was reduced. For $Y_p \leq .239$ no range of $\eta_{10}$ was allowed when this constraint was combined with the D + $^3$He bound. Previously we derived a lower bound on $Y_p$ of .238 if D+$^3$He was used alone to first bound $\eta_{10}$, and then the $\eta_{10}$ value was used to bound $Y_p$ (to compare to earlier such bounds (i.e. [11]). The new neutron half life would not change that bound. However in any case the newly derived bound of .239 obtained using the correlated constraints is more stringent, and more consistent. If the primordial helium abundance is determined empirically to be less than this value with great confidence, and the D + $^3$He upper limit remains stable, standard BBN would be inconsistent with observation.

Table 4
$^4$He, D and $^7$Li Estimates & $\eta_{10}$ limits ($^7$Li$_{-10}$ <5; $D_{-5}$ > 2)

| $Y_{p_{max}}$ | $\eta_{10_{max}}$ | # allowed models: {$^4$He & D & $^7$Li} ($^4$He & D:$^4$He & $^7$Li:D & $^7$Li) ($^4$He:D:$^7$Li) |
|---|---|---|
| .248 | 6.94 | 48 (136:53:156) (178:516:203) |
| .249 | 7.22 | 52 (177:101:64) (654:217:136) |
| .250 | 7.24 | 47 (191:113:47) (995:191:113) |

## 5. CONCLUSIONS AND CHALLENGES

There can be little doubt that the present ability of BBN to constrain cosmological parameters is almost completely governed by systematic uncertainties in our inferences of the actual light element primordial abundances. Nevertheless, the fact that such systematic uncertainties need not be gaussian does not block our ability to utilize the statistically meaningful uncertainties in BBN predictions. As long as we are willing to quote conservative one-sided limits on the various abundances which incorporate reasonable estimates of the systematic uncertainties then the determination of what confidence levels can be assigned to various theoretical predictions is straightforward. Moreover, as the observational limits on various elemental abundances is varied, the significance of the different elements for constraining cosmological parameters varies. In addition, for a non-trivial range in $\eta_{10}$, correlations exist between the various abundance predictions, and a self consistent use of all available constraints is important. Finally, $Y_p$, in spite of its systematic uncertainty, plays a dominant role unless one is willing to accept an upper limit of greater than .247. Beyond that, the convergence of D and $^7$Li limits suggest a safe upper bound of on the baryon density today of less than 16% of closure density.

We thus find ourselves with on the verge of several possible interesting inconsistencies:
1. If the primordial $^4He$ fraction can definitively be established to be less than 23.9% either D +$^3$He estimates will have to be revised, or some more dramatic cosmological or particle physics-based alteration in BBN predictions will be required
2. Recent estimates of the baryon fraction in rich clusters (i.e. [49]) suggest that this fraction can be rather large. If the Universe is flat, then the baryon fraction which one would derive based on the rich cluster estimates is at least a fraction of 2 larger than that allowed by our BBN estimates, even with systematic uncertainties allowed for. Whether this is an indication that the universe is not flat, or an indication that the cluster estimates themselves suffer from large possible systematic uncertainties remains to be seen

In any case, as time proceeds and more independent observations are made we will undoubtedly get a better handle on the systematic uncertainties which presently limit the efficacy of BBN constraints. As I have shown here, several very interesting possible constraints on cosmology and particle physics are around the corner. Until then, BBN analyses still provide among the most useful tools to constrain fundamental physics. The updated tables and relations presented here should provide a useful reference to allow researchers to translate their own limits on the light element abundances into meaningful bounds on $N_\nu$ and $\eta_{10}$.

I would like to thank my present BBN collaborator Peter Kernan, for his significant contributions to the work discussed here